\documentclass[journal]{IEEEtran}
\usepackage{latexsym}
\usepackage{float}
\usepackage{amsfonts}
\usepackage{amsbsy}
\usepackage{amssymb}
\usepackage{times}
\usepackage{graphicx}
\usepackage{setspace}
\usepackage{enumerate}
\usepackage[usenames]{color}
\usepackage[dvips]{pstcol}
\usepackage{epstopdf}
\usepackage[caption=false]{subfig}
\usepackage{cite}
\usepackage{amssymb}
\usepackage{amsfonts}
\usepackage{graphicx}
\usepackage{epsfig}
\usepackage{psfrag}
\usepackage{xcolor}
\usepackage{amsfonts, bm}
\usepackage{epstopdf}
\usepackage{cite}
\usepackage{color}
\usepackage{xcolor}
\usepackage{subfig}
\usepackage{verbatim}
\usepackage{multirow}
\usepackage{booktabs}
\usepackage{amsthm}
\usepackage{makecell}
\usepackage{units}
\usepackage[linesnumbered, ruled]{algorithm2e}
\usepackage{algpseudocode}
\usepackage{amsmath}

\IEEEoverridecommandlockouts

\begin{document}

\title{Towards Compatible Semantic Communication: A Perspective on Digital Coding and Modulation}

\author{ Guangyi Zhang, Kequan Zhou, Yunlong Cai, Qiyu Hu, and Guanding Yu
\thanks{ © 2024 IEEE. Personal use of this material is permitted. Permission from IEEE must be obtained for all other uses, in any current or future media, including reprinting/republishing this material for advertising or promotional purposes, creating new collective works, for resale or redistribution to servers or lists, or reuse of any copyrighted component of this work in other works.
 } } 


\maketitle

\vspace{-1.3em}


\IEEEpeerreviewmaketitle

\begin{abstract}
Semantic communication (SC) is emerging as a pivotal innovation within the 6G framework, aimed at enabling more intelligent transmission. This development has led to numerous studies focused on designing advanced systems through powerful deep learning techniques. Nevertheless, many of these approaches envision an analog transmission manner by formulating the transmitted signals as continuous-valued semantic representation vectors, limiting their compatibility with existing digital systems. To enhance compatibility, it is essential to explore digitized SC systems. This article systematically identifies two promising paradigms for designing digital SC: probabilistic and deterministic approaches, according to the modulation strategies.  For both, we first provide a comprehensive analysis of the methodologies. Then, we put forward the principles of designing digital SC systems with a specific focus on informativeness and robustness of semantic representations to enhance performance, along with constellation design. Additionally, we present a case study to demonstrate the effectiveness of these methods. Moreover, this article also explores the intrinsic advantages and opportunities provided by digital SC systems, and then outlines several potential research directions for future investigation.
\end{abstract}


\section{Introduction}
Wireless communication has advanced to the 5G era, with the next generation envisioned as a more intelligent system supporting  diverse applications, such as the metaverse and extended reality.
These services, predominantly driven by advanced artificial intelligence (AI) techniques, demand high reliability and massive data traffic.
To address these challenges, semantic communication (SC) has emerged as a pivotal mechanism for next-generation networks, focusing on transmitting only task-relevant information to improve efficiency and reduce transmission overhead.

\begin{figure*}
	\begin{centering}	
		\includegraphics[width=0.68\textwidth]{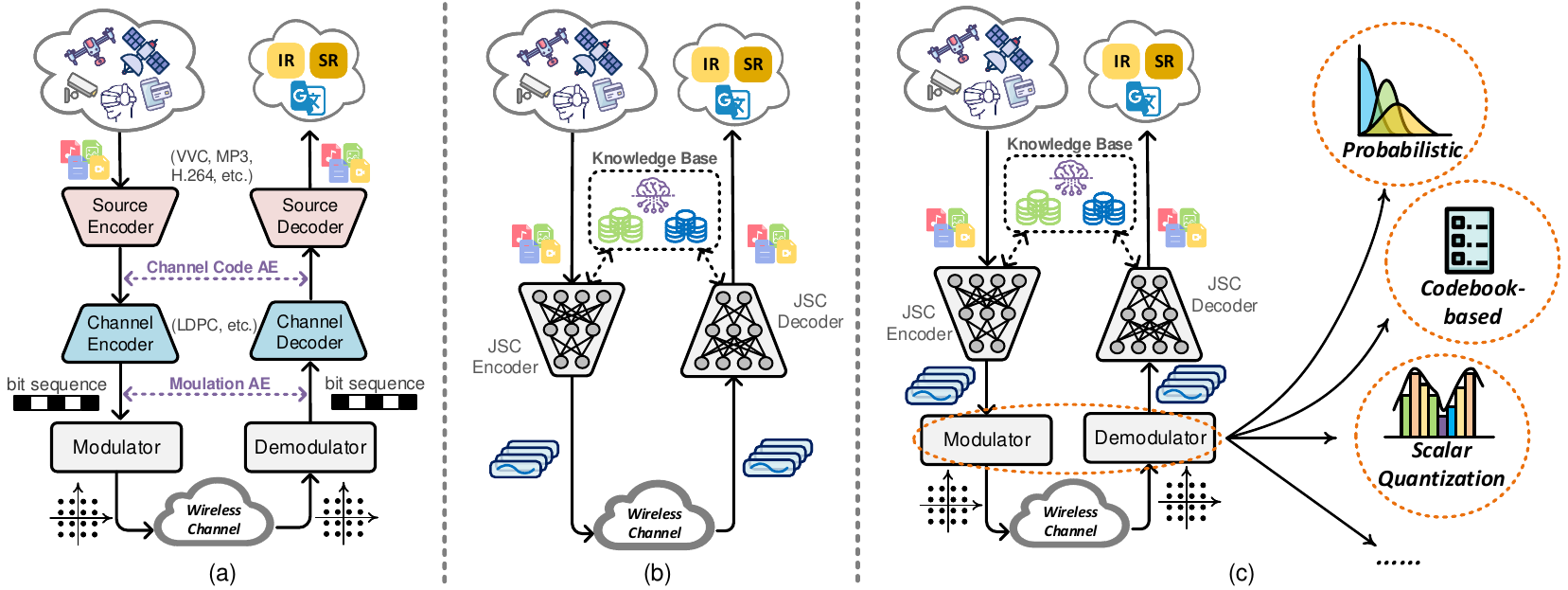}
		\par \end{centering}
	\caption{The illustration of the paradigm shift of SC systems. (a) Separation-based systems denote the conventional scheme that exploits a combination of source codecs (VVC, MP3, H.264, etc.); (b) Analog SC is firstly proposed for the transmission of various modalities of data, where the channel symbols are assumed to be analog; (c) Digital SC  conceive of designing SC systems by separating the modulation module again for better compatibility, aligning with the physical layer techniques.}
	\label{ParadigmShift}
\end{figure*}

In recent years, deep learning (DL) has shown substantial potential in shaping modern SC frameworks. The paradigm for designing SC systems involves leveraging deep neural networks to achieve joint source-channel coding (JSCC). This approach aims to enhance task performance and error resilience by capturing compact and robust semantic representations. 
Focusing on this objective, many studies have utilized advanced neural network (NN) models (e.g., Transformer and Mamba) in SC systems to directly generate semantic representations, which serve as channel input symbols.  The semantic receiver then executes the task  leveraging another NN models, usually trained in a data-driven manner. 
Current research primarily focuses on analog designs for SC, where semantic representations are transmitted as continuous signals over wireless channels. This approach necessitates infinite-precision constellations or analog modulation, exhibiting low compatibility with existing digital communication system regarding hardware and protocols. To overcome these challenges, there is a clear need to develop digitized SC systems for better compatibility with real-world hardware. Among different potential countermeasures, the coding and modulation are of great significance.  To ensure SC systems integrate smoothly with existing digital infrastructure, it is essential to investigate how semantic information can be effectively represented and transmitted using a finite set of symbols \cite{Mingze_DMA}.  Consequently, one of the key challenges in SC can be articulated as: \cite{Huiqiang2024Towards}: \textit{How can efficient discrete semantic representation be obtained using deep learning methods?}

To explore the answer, we can take a glance at discrete neural representation learning (NRL) \cite{pmlrST}, which is closely related due to its emphasis on the informativeness of discrete representations for tasks such as image compression and generative modeling. However, beyond informativeness, SC systems also face a critical challenge in ensuring \textit{robustness} against channel noise, as semantic representations are inevitably subject to distortion during transmission.
In this context, designing effective digital SC systems requires a focus on both the properties by improving semantic extraction (coding) and representation discretization (modulation). 
This leads to two key areas for advancing digital coding and modulation in SC systems. First, in terms of coding, advanced model architectures and effective training strategies, such as unsupervised pretraining, are crucial for learning semantic representations that are both informative and robust. Second, existing discretization techniques in discrete NRL, such as vector quantization (VQ) and scalar quantization (SQ), show promise for enhancing modulation efficiency but also necessitate improvements for maintaining robustness against noise.

This article investigates digital SC systems from  the perspective of digital coding and modulation. After reviewing current designs, we present two promising paradigms for designing digital SC systems: the probabilistic and deterministic approaches. The probabilistic approach employs probabilistic modeling, where the encoding process is stochastic. In this case, the channel input symbols are generated by sampling from a probability distribution conditioned on the input data. The deterministic approach, on the other hand, constructs a direct mapping from continuous semantic representations to discrete symbols.
For both approaches, we provide an in-depth analysis of their methodologies and the faced challenges. Motivated by the success in NRL, we also propose key principles for designing digital SC systems, encompassing a general procedure.
We predominantly concentrate on the system formulation, multiphase training strategy, alongside the constellation design. Additionally, we offer a case study to demonstrate the effectiveness of these designs and highlight some points for further improvement. Furthermore, this article explores the inherent advantages and opportunities of adopting digital SC systems. Finally, we outline open issues and potential research directions before concluding. In summary, this article serves as a tutorial on the implementation of a digital SC system from the perspective of coding and modulation, and aims to inspire further research in this rapidly evolving field.

\section{Paradigm Shift}
In this section, we concentrate on introducing the evolution of SC designs from a fundamental viewpoint of coding and modulation. We begin by reviewing traditional data transmission methods, then discuss key approaches for implementing analog SC systems, and finally provide an overview of digital SC designs.

The rapid advancement of wireless networks and edge devices has led to an explosion of data, which is critical for AI services that require the integration of multiple data types to perform complex tasks. As illustrated in Fig. \ref{ParadigmShift}(a), a two-step coding strategy is typically employed for data transmission: source coding for compression, followed by channel coding for error correction, modulation, and mapping to antenna ports. Moreover, NN models have been widely employed to improve the performance of each procedure, with notable approaches including modulation autoencoders that learn the modulator-demodulator process and channel coding autoencoders that jointly optimize channel coding and modulation.

In contrast, SC is initially designed using a JSCC approach, which further integrates the source coding, channel coding, and modulation. In this framework, continuous semantic representations are treated directly as channel symbols. As illustrated in Fig. \ref{ParadigmShift}(b),  a typical analog SC system comprises a joint source-channel (JSC) encoder, a knowledge base, a JSC decoder, and a task component. 

\begin{figure*}
	\begin{centering}	
		\includegraphics[width=0.62 \textwidth]{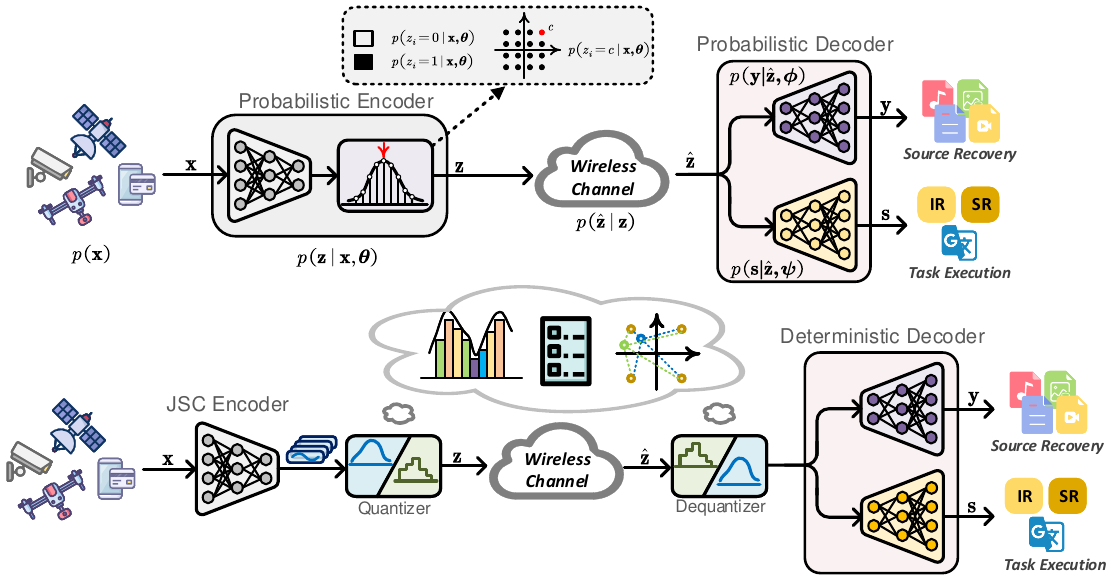}
		\par \end{centering}
	\caption{The illustrations of the probabilistic method and deterministic method. }
	\label{ProDeter}
\end{figure*}

To ensure compatibility with existing digital systems, the coding and modulation steps can be technically separated. Consequently, as shown in Fig. \ref{ParadigmShift}(c), a digital SC can be formulated as a weakly-coupled joint coding and modulation scheme:
\begin{itemize}
	\item \textit{Knowledge Base:} The knowledge base is a well-established context that encompasses information related to the source, model, task, and channel. It is typically defined as a model with processing, reasoning, and memory capabilities.
		
	\item \textit{JSC Encoder}: The encoder maps the source data to a semantic representation, incorporating task-specific information. It implicitly performs both semantic encoding and channel coding, as it is jointly trained to account for channel noise.
	
	\item \textit{Modulator}: A separate modulation module is designed to convert continuous representation to discrete representation. This discrete representation can be  finite constellation symbols as in traditional systems. 
	
	\item \textit{Wireless Channel:} This presents a challenge for the encoder in generating a robust semantic representation. Typically, specific transmission techniques, such as layer mapping and beamforming, are inherently integrated within the channel transfer function.
	
	\item \textit{JSC Decoder:} The JSC decoder performs both channel decoding and source decoding. Utilizing the task information, it generates the desired results accordingly.
\end{itemize}

In the next section, we elaborate on the implementation of joint coding and modulation, highlighting two key approaches: probabilistic and deterministic methods.

\section{Digital SC: Methodologies Overview} \label{MethodologyChallenges}
In this section, we introduce the methodologies of existing digital SCs and highlight the challenges to address.
In general, the methods for implementing digital  SC systems can be categorized into two groups: (i) probabilistic modulation methods, which utilize probabilistic networks; and (ii) deterministic modulation methods, which involve direct quantization.
\subsection{Probabilistic Methods}

As illustrated in Fig. \ref{ProDeter}, probabilistic modulation methods employ a probabilistic encoder-decoder pair to construct digital SC systems.
These techniques  aim to learn the transition probability from source data to digital bits or constellation symbols via a probabilistic network. Then, the exact bits or constellation symbols are generated by sampling from the produced distribution.

One branch of probabilistic modulation methods focuses on learning the transition from source data to channel-input bit sequences,  particularly for discrete channel models such as the binary erasure channel (BEC) and binary symmetric channel (BSC) \cite{Choi2019neural}. 
 In these approaches, the encoder acts as a Bernoulli random variable generator and its distribution parameters are determined by NN models.
The encoder probabilistically generates the channel-input bit sequence based on the source data.
Conversely, the decoder, also parameterized by NN models, reconstructs the source data from the received noisy bit sequence.
By considering the transition characteristics of the channel model, the entire process can be viewed as a Markov chain, as shown in Fig. \ref{ProDeter}.  
The objective is to maximize the mutual information between the source data and the noisy bit or symbol sequence.
Moreover, transition probabilities from source data to constellation symbols can also be learned using an NN-based probabilistic encoder-decoder architecture \cite{Bo2024joint}, inspired by the  variational autoencoder (VAE) framework.
In this setup, the encoder generates constellation symbols probabilistically, while the decoder recovers the source data from the received signal.
The aim for the encoder is to retain maximal information about the source data within the received signal, while  the decoder ensures that the  NN-parameterized posterior distributions converge towards the true posterior distributions.

However, the aforementioned approaches encounter several common challenges,  including issues related to intractable posterior probability inference, and gradient estimation.
Firstly, exact inference for these probability distributions in Bayesian models is computationally intractable.
A feasible solution to this issue involves utilizing variational inference techniques, which provide a tractable lower-bound approximation of the posterior probability \cite{Blei2017variational}.
Another significant challenge is gradient estimation, which arises when optimizing the parameters of the encoder and decoder.
To address this problem, solutions include estimating the gradients utilizing the score function estimator\cite{Bo2024joint} and employing the reparameterization trick commonly used in VAEs to reduce estimation variance.
Continuous relaxation methods, like Gumbel-Softmax, can also help \cite{Zhang2024a},  although they may lead to performance degradation due to mismatches between training and testing.
Another effective solution involves using a multisample variational lower bound to achieve low-variance gradients \cite{Choi2019neural}.

\subsection{Deterministic Methods}
As presented in  Fig. \ref{ProDeter}, deterministic modulation methods  adapt traditional analog SC systems by integrating a quantizer-dequantizer pair into their frameworks. 
The quantization process differs according to specific objectives, which generally fall into three categories: scalar quantization, symbol quantization, and vector quantization.

\textbf{Scalar Quantization:}
	Scalar quantization methods target the floating-point scalar elements within the encoded semantic features for quantization \cite{Chuanhong2024}.
	These elements are quantized into integers, converted into bit sequences, and processed using conventional digital modulation techniques.  This type of quantization has been widely employed in realm of network quantization for its potential to minimize performance loss associated with parameter quantization.
	
\textbf{Symbol Quantization:}
	In analog SC systems, the encoded feature vectors are generally mapped to constellation symbols.
	The symbol quantization methods just aim to remap these irregular constellation symbols into a predefined symbol space, which typically corresponds to modern digital modulation schemes like M-QAM. Intuitively, symbol quantization can be viewed as a special case when the modulation order equals to the  number of quantization levels in scalar quantization, simplifying the system formulation. 
	
\textbf{Vector Quantization:}
	The vector quantization methods focus on designing a discrete codebook for the semantic feature space, quantizing each encoded feature vector into the nearest basis vector in this codebook.
	These basis vectors are trainable parameters that are optimized alongside the parameters of the encoder and decoder.
	By sharing the codebook between the transmitter and receiver, encoded feature vectors can be represented by the indices of  these basis vectors, significantly reducing the transmission overhead.
	These indices are then be converted into bits and transmitted using digital modulation techniques. Notable examples include VQ-VAE2 and residual vector quantizers (RVQ), which have been widely used in neural speech coding designs.

Despite different discretization objectives, these methods share a core strategy: using differentiable approximations for quantization to address non-differentiability issues. Four widely adopted differentiable approximation techniques are NN approximation, straight-through estimator, soft-to-hard annealing, and additive uniform noise. In the NN approximation approach, the quantizer and dequantizer are replaced with neural network-based counterparts \cite{Jiang2022deep}.
Here, the NN quantizer is implemented through a fully connected layer that transforms floating-point semantic features into a bit sequence.
The straight-through estimator (STE) enables gradients to pass directly from the decoder input to the encoder output, bypassing the quantization layer to facilitate backpropagation.
The soft-to-hard annealing method approximates the quantization with a controllable soft relaxation managed by a temperature parameter that gradually decreases  during training.
As the temperature approaches zero, this soft relaxation  effectively converges to the actual quantization process, addressing the non-differentiability issue while mitigating the mismatch between training and testing phases.
In \cite{Balle2017endtoend}, the additive uniform noise is introduced into the quantization objective during training, replacing the direct quantization to enable smooth end-to-end optimization.

\begin{figure*}
	\begin{centering}	
		\includegraphics[width=0.62  \textwidth]{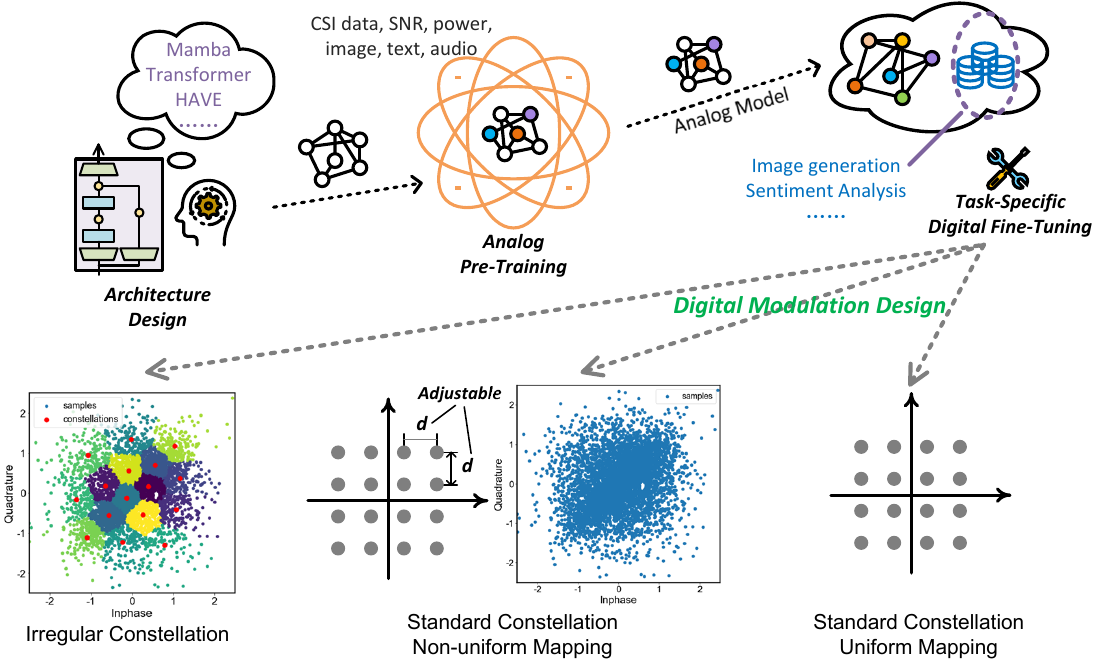}
		\par \end{centering}
	\caption{The illustration of the proposed multistage design strategy. For the constellation illustration, we view half of the semantic representation output by a deterministic modulator as the real part, while the left half is regarded as the imaginary part. }
	\label{multistage}
\end{figure*}

\section{Principles for High-Efficiency Digital SC}
In this section, we present the detailed design of high-efficiency digital SC systems and introduce a mutual-stage design strategy. 

\subsection{Challenges}
While the methods mentioned above help enable digital SC systems, several challenges remain.
\begin{itemize}
	\item \textbf{Challenge 1:} Most probabilistic methods  rely on variational formulations, which have a similar structure to the VAE model. However, the VAE model has been shown to scale poorly with large-scale source information, such as high-resolution images.
	
	\item \textbf{Challenge 2:} Although several advanced neural network architectures have been developed for implementing digital SC, methods to enhance their effectiveness remain in the early stages, underscoring an urgent need for advanced learning strategies.
	
	\item \textbf{Challenge 3:} Previous works  primarily utilize standard constellations for semantic representation, which may be suboptimal given the non-uniform characteristics of the representation.
\end{itemize}
To address these challenges, we develop a structured design process for achieving an efficient digital SC system, with proposed solutions at each step.

\subsection{Key Designs: Multistage Design Strategy}
\subsubsection{System Formulation and Architecture Design}
	Probabilistic  methods, such as standard VAEs and amortized models,  incorporate hierarchical stochastic latent variables with high expressive capacity, yet face training challenges in deep hierarchies due to multiple layers of conditional dependencies.  Consequently,  digitized SC systems based on VAEs have often been constrained to shallow models, leading to limited representation capability for high-dimensional data.
	A promising approach involves employing autoregressive models, including autoregressive flows and hierarchical VAEs (e.g., hierarchical VQ-VAE, ResNet VAE, and Nouveau VAE), which have demonstrated effectiveness in generating high-quality data across diverse modalities.
	
	To obtain expressive semantic representations, it is essential to leverage powerful NN architectures such as Transformers and Mamba. Transformers excel at processing sequential data and capturing long-range dependencies, while Mamba, a recently introduced architecture, has demonstrated superior performance and inference speed compared to Transformer-based models. By utilizing various carefully designed operations, these architectures can effectively capture subtle semantic differences, resulting in more accurate information transmission.

\subsubsection{Training for Compact Semantic Representation}
	While a well-designed model can facilitate the extraction of semantic representations,  training such a model often presents challenges. The learning process of a digital SC system can intuitively be viewed as a problem of NRL \cite{pmlrST}, specifically discrete NRL. This process can be classified as either supervised or unsupervised (self-supervised), depending on the use of labeled data during training. These approaches can be integrated into digital SC system design by incorporating regularization terms into the training objective or by following specific principles, such as the information bottleneck and maximal coding rate reduction. Additionally, pre-training strategies have proven to be highly effective in enhancing learning performance.

	The pre-training strategy has revolutionized  DL,  particularly with the advent of large language models like GPT family.  During the pre-training phase, models learn from extensive datasets,  such as CSI data and images, to develop robust semantic representations in an unsupervised or self-supervised manner.  Subsequently, these models can then be fine-tuned on smaller, task-specific datasets to realize high performance across various applications, from text generation to sentiment analysis.
	Recent advancements in NRL suggest that training in the analog domain enhances feature extraction. Building on this insight, we propose treating the development of digital systems as a downstream  task, suggesting that pre-training occurs in the analog domain, with the learned parameters then transferred to the digital domain for fine-tuning. This approach offers two key advantages:
	\begin{itemize}
		\item The mechanisms for designing analog SC systems can be effectively transferred to digital systems.
		\item It facilitates the development of a robust encoder that produces expressive semantic representations, which is crucial for building an efficient digital system.
	\end{itemize}

\subsubsection{Fine-Tuning for Digital SC}
Fine-tuning is the phase that follows pre-training,  during which the model is adapted to a specific task or dataset. This process refines the model's parameters by training it on a task-specific dataset,  enabling  it to leverage the generalized knowledge acquired during pre-training for improved performance on specialized tasks. In our context, we fine-tune the SC system using digital modulation during this phase, utilizing any scheme discussed in Section III. Given the characteristics of the representation space obtained during pre-training, we propose three strategies for designing modulators to address Challenge 3.

\textbf{Standard Constellation with Uniform Mapping:} The most straightforward approach to implementing a digital SC system is to map continuous semantic representations into discrete symbols using a standard constellation. For deterministic modulation, this is accomplished by applying a quantization-like operation to each element in the semantic representation vector. In the case of a probabilistic modulator, a standard constellation serves as the discrete sampled points. This method effectively simulates transmission errors, allowing for end-to-end training that enhances the system's robustness.

\textbf{Standard Constellation with Non-Uniform Mapping:} 
A standard constellation achieves satisfactory performance when input symbols are uniformly distributed. However, our analysis of the output from an analog SC system reveals that this condition is not met due to the non-uniform nature of the model output. As shown in Fig. 3, the elements of the semantic representation are not uniformly distributed. Consequently, uniformly modulating the representation may not be appropriate. To address this mismatch, employing a flexible modulator is crucial. One approach is to use a learnable parameter to define an appropriate distance between two constellations. Alternatively, one can manually establish suitable decision boundaries instead of relying solely on nearest-neighbor mapping. In \cite{ParkDigital}, the authors derived closed-form decision boundaries that significantly enhance demodulation efficiency, resulting in improved task accuracy, particularly in low SNR conditions.

\textbf{Irregular Constellation:}
To better accommodate the non-uniform nature of analog semantic representations, an effective solution is to design a corresponding irregular constellation.  This can be realized by conducting clustering techniques to the continuous semantic representation, using unsupervised algorithms such as K-means or mixed Gaussian models. Fine-tuning digital SC systems with the resulting irregular constellations can yield more compatible semantic representations \cite{Mengyang_SPL}.  Additionally, if we allow the cluster constellation points to be learnable during the fine-tuning phase, the model can iteratively refine and optimize the constellation for digital representation. In the case of a probabilistic design, the constellation can also be dynamically adjusted during training based on the predicted probability for each constellation symbol.
\begin{figure}
	\begin{centering}	
		\includegraphics[width=0.4 \textwidth]{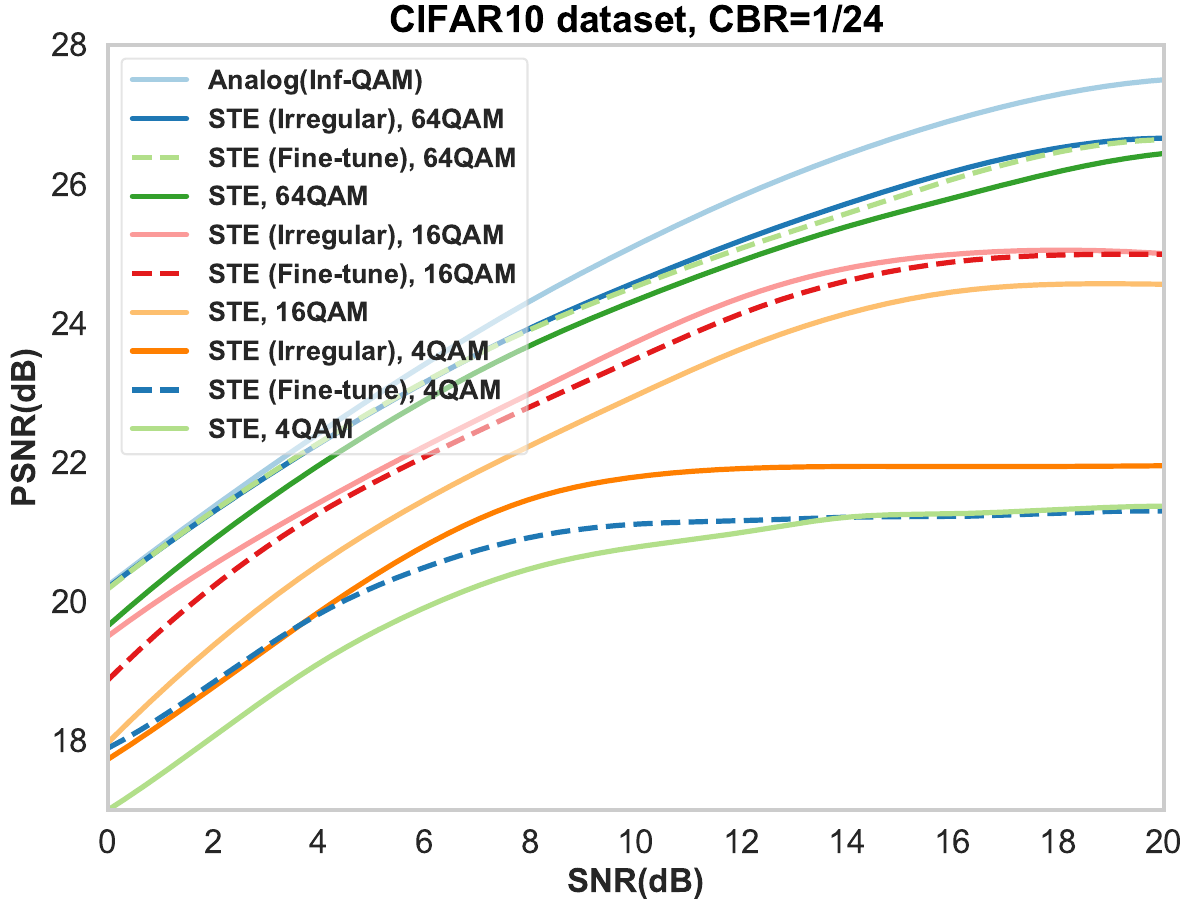}
		\par \end{centering}
	\caption{The performance of different schemes with different modulation orders versus SNR. }
	\label{psnrfigure}
\end{figure}
\subsection{Case Study}
We provide numerical simulations to verify the performance of the proposed strategies. The experiments are conducted over additive white Gaussion noise (AWGN) channels using the classical CIFAR-10 dataset. We focus on image transmission in these experiments, as it has garnered significant attention in the research community. To ensure fairness, all benchmark models share the same architecture and training settings. The details of the considered schemes are as follows:
\begin{itemize}
	\item  ``Analog": This approach involves training the model in an analog manner, utilizing a full-resolution constellation.
	\item  ``STE" : In this scheme, the model is trained directly for the digital SC system, where the STE is employed to address the gradient propagation.
	
	\item  ``STE (Fine-tune)": In this approach, the model is fine-tuned for digital transmission using parameters that were pre-trained in an analog system.
	
	\item  ``STE (Irregular)": In this approach, the model is fine-tuned for digital transmission using parameters pre-trained in an analog system, where the constellation points are generated using the K-means clustering method.
\end{itemize}
In Fig. \ref{psnrfigure}, we compare the performance of  various schemes across different SNR regimes,  focusing on three modulation types. Our key observations are as follows: i)The performance of digital SC improves with the modulation orders, and thus higher modulation order is generally preferable. This is quite different from conventional systems, where lower modulation orders typically achieve better bit error rates, even at the same transmission rate; ii) Comparing STE with STE (Fine-tuning) reveals that the pre-training phase greatly enhances the network's expressive capability, allowing it to produce informative and robust discrete semantic representations. iii) The scheme fine-tuned with irregular constellations outperforms all other digital schemes, while the scheme fine-tuned with regular constellations comes in a close second. This clearly demonstrates the effectiveness of the proposed methods.

\begin{figure}
	\begin{centering}	
		\includegraphics[width=0.48 \textwidth]{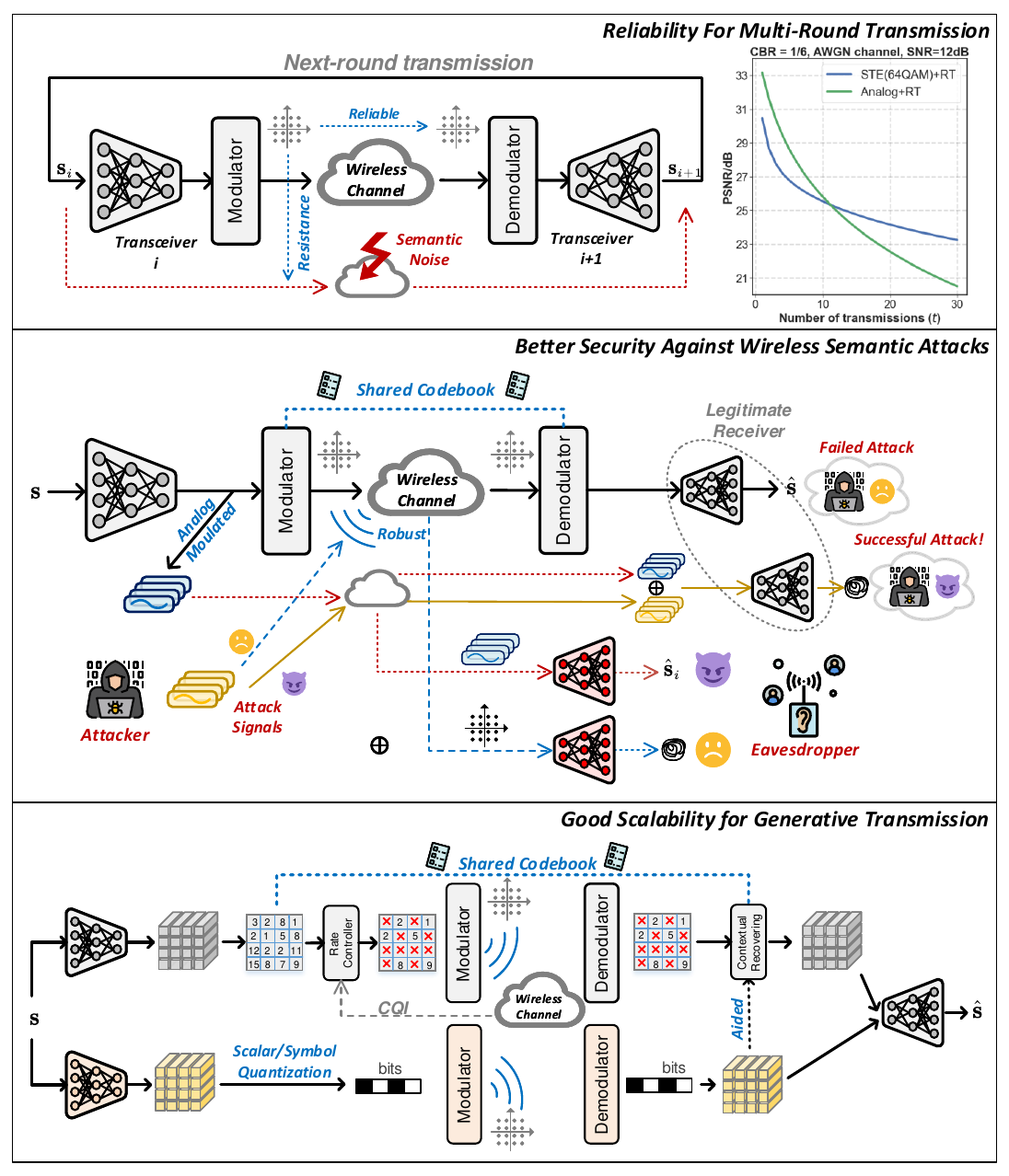}
		\par \end{centering}
	\caption{Three advantages of digital SC system for further investigation. The top part displays the reliability of multiround transmission. The medium illustrates the robustness against wireless attacks. The bottom illustrates the proposed hybrid architecture for generative transmission.}
	\label{advan}
\end{figure}

\section{Future Research Directions}
In this section, we first outline several directions for boosting the inherent advantages of digital SC systems. Then, several open challenges are further discussed. 

\subsection{Inherent Advantages of Digital SC}
\subsubsection{Reliability For Multiround Transmission}
In \cite{GuangyiICL}, we investigated a crucial  issue of distortion accumulation in deep learning-based SC systems.  Most existing SC approaches rely on lossy transmission, leading to distortion that accumulates as the receiver further transmits images to another device.   Generally, digital systems demonstrate stronger resistance to interference compared to analog systems; digital signals have a limited number of discrete values, indicating that noise does not directly impact each value when disturbance is small. While noise can cause slight variations, as long as these do not exceed the system's threshold, the signal remains intact. Inspired by this, digital SC systems are anticipated to offer significantly better resistance to distortion accumulation. Our initial results, illustrated in Fig. \ref{advan}, confirm the advantages of digital SC systems in addressing this issue.

\subsubsection{Security Against Wireless Semantic Attacks}
While extracted semantic representations contain task-related information and offer a degree of privacy, they remain vulnerable to malicious attacks and eavesdropping. The open nature of wireless systems makes SC systems particularly susceptible to adversarial threats. Additionally, neural networks face the risk of information leakage; an eavesdropper could use model inversion techniques to reconstruct raw data from the received semantic representations. Digital design presents a promising avenue for increasing the security of SC systems. Firstly, VQ approaches involve discretely mapping  continuous semantic representations using a codebook, inherently reducing the amount of detailed information that can be inferred from the decoded symbol (i.e., codebook index). Secondly, VQ demonstrates robustness against adversarial attacks \cite{Sagduyu_IMG}. Thirdly, integrating bit conversion into the SC system allows for the incorporation of traditional encryption algorithms, further improving security and compatibility.

\subsubsection{Good Scalability for Generative Transmission}
In the realm of neural image compression,  perceptual compression—often referred to as generative compression—has recently garnered significant attention.  The main idea is to leverage generative models, such as generative adversarial networks (GANs), VAEs, and diffusion models, to compress images into bit streams, aiming to achieve better detail and perceptual quality at extremely low bit rates. 
In this low bit-rate region, the VQ-based methods have proven to be the dominant techniques, primarily due to their ability to capture correlations among different latent dimensions \cite{JinchengIWC}. Inspired by this, generative SC systems can prioritize using the VQ-based paradigms, potentially integrating forward error correction coding to further enhance performance. While VQ methods encounter challenges at higher bit rates, a hybrid scheme combining SQ and VQ, as illustrated in Fig. \ref{advan}, shows significant potential.

\subsection{Open Challenges}
\subsubsection{End-to-End Joint Optimization}
While JSCC-based digital SC systems have notable advantages, their  tightly coupled design can pose implementation challenges, particularly in cross-layer design. In this context, exploring weakly coupled approaches that combine NN-based compression models with efficient error correction codes (e.g., polar codes and channel coding autoencoder) is crucial for protecting encoded features from channel distortion. This involves assessing the semantic importance of extracted features to help the channel encoder prioritize critical elements for unequal error protection. Additionally, using different pre-trained NNs for source coding and channel encoding based on the specific wireless environment is also a viable strategy.

\subsubsection{Implementation on Hardware Platforms}
A successful implementation of digital SC relies not only on the design of neural network architectures and training strategies, but also on efficient realization across hardware platforms.
Firstly, the energy and latency overhead of digital SC is highly dependent on the hardware.
Specialized hardware should optimize power consumption and enhance real-time performance, ensuring  effective operation in energy-constrained environments and applications that demand quick responses.
Secondly, hardware implementations should also be customizable to meet the specific requirements of different applications.

\subsubsection{Adaptive Coding-Modulation Schemes}
In digital SC systems,  relying on a fixed coding-modulation scheme can result in performance degradation as channel conditions fluctuate.
To mitigate  this issue, an adaptive coding-modulation scheme is essential for enhancing transmission reliability and optimizing spectral efficiency.
This approach involves estimating the CSI at the receiver, feeding the estimation back to the transmitter, and dynamically adjusting the coding-modulation strategies.
Key adjustable parameters include data rate, modulation order, and constellation patterns, where a unified and multifunctional digital SC system is of significance.

\section{Conclusion}
In this article, we focused on the coding and modulation challenges involved in realizing digital SC systems. We identified two typical paradigms and provided a detailed analysis of their methodologies and challenges. To guide the design of digital SC systems, we proposed a multiphase strategy that emphasizes the system's representation capability and robustness. Specifically, we suggested viewing the digitization process as a downstream task and highlighted three types of constellations for fine-tuning. Some important directions for enhancing the inherent advantages as well as overcoming the challenges were also highlighted. This article provides insightful guidance for future research on digital SC systems.

\bibliographystyle{IEEEtran}
\bibliography{IEEEabrv,Reference}
\vspace{-.4in}
\begin{IEEEbiographynophoto}
	{Guangyi Zhang} [S] (zhangguangyi@zju.edu.cn) is currently pursuing the Ph.D. degree with the College of Information Science and Electronic Engineering, Zhejiang University.
\end{IEEEbiographynophoto}	
\vspace{-.45in}
\begin{IEEEbiographynophoto}
	{Kequan Zhou} [S] (kqzhou@zju.edu.cn) is currently pursuing the Ph.D. degree with the College of Information Science and Electronic Engineering, Zhejiang University.
\end{IEEEbiographynophoto}
\vspace{-.45in}
\begin{IEEEbiographynophoto}
	{Yunlong Cai} [SM] (ylcai@zju.edu.cn) is a Professor with the College of Information Science and Electronic Engineering, Zhejiang University, Hangzhou, China.
\end{IEEEbiographynophoto}
\vspace{-.45in}
\begin{IEEEbiographynophoto}
	{Qiyu Hu} [S] (qiyhu@zju.edu.cn) is with Zhejiang Provincial Government, Hangzhou, China.
\end{IEEEbiographynophoto}
\vspace{-.45in}
\begin{IEEEbiographynophoto}
	{Guanding Yu} [SM] (yuguanding@zju.edu.cn) is a Professor with the College of Information Science and Electronic Engineering, Zhejiang University, Hangzhou, China.\\
\end{IEEEbiographynophoto}

\end{document}